# Metal-to-insulator transition in oxide semimetals by anion doping

**Authors**: Haitao Hong,[1,2,†] Huimin Zhang,[3,†] Shan Lin,[4] Jeffrey A. Dhas,[5,6] Binod Paudel,[5] Shuai Xu,[1,2] Shengru Chen,[1,2] Ting Cui,[1,2] Yiyan Fan,[1,7] Dongke Rong,[1] Qiao Jin,[1] Zihua Zhu,[8] Yingge Du,[5] Scott A. Chambers,[5] Chen Ge,[1,2] Can Wang,[1,2,9] Qinghua Zhang,[1] Kui-juan Jin,[1,2,9,*] Le Wang,[5,*] Shuai Dong,[3,*] and Er-Jia Guo[1,2,*]

**Affiliations**:

[1] Beijing National Laboratory for Condensed Matter Physics and Institute of Physics, Chinese Academy of Sciences, Beijing 100190, China

[2] Department of Physics & Center of Materials Science and Optoelectronics Engineering, University of Chinese Academy of Sciences, Beijing 100049, China

[3] Key Laboratory of Quantum Materials and Devices of Ministry of Education, School of Physics, Southeast University, Nanjing, 211189, China

[4] Materials Science and Technology Division, Oak Ridge National Laboratory, Oak Ridge, TN, 37831, USA

[5] Physical and Computational Sciences Directorate, Pacific Northwest National Laboratory, Richland, WA 99354, USA

[6] School of Chemical, Biological and Environmental Engineering, Oregon State University, Corvallis, OR, 97331, USA

[7] Beijing Advanced Innovation Center for Materials Genome Engineering, Department of Physical Chemistry, University of Science and Technology Beijing, Beijing 100083, China

[8] Environmental Molecular Sciences Laboratory, Pacific Northwest National Laboratory, Richland, WA 99354, USA

[9] Songshan Lake Materials Laboratory, Dongguan, Guangdong 523808, China

† These authors contribute equally to the manuscript.

E-mail: kjjin@iphy.ac.cn, le.wang@pnnl.gov, sdong@seu.edu.cn, and ejguo@iphy.ac.cn

**Abstract**

Oxide semimetals exhibiting both nontrivial topological characteristics stand as exemplary parent compounds and multiple degrees of freedom, offering great promise for the realization of novel electronic states. In this study, we present compelling evidence of profound structural and transport phase shifts in a recently uncovered oxide semimetal, $SrNbO_3$, achieved through effective *in-situ* anion doping. Notably, a remarkable increase in resistivity of more than three orders of magnitude at room temperature is observed upon nitrogen-doping. The extent of electronic modulation in $SrNbO_3$ is strongly correlated with the misfit strain, underscoring its phase instability to both chemical doping and crystallographic symmetry variations. Using first-principles calculations, we discern that elevating the level of nitrogen doping induces an upward shift in the conductive bands of $SrNbO_{3-\delta}N_\delta$. Consequently, a transition from a metallic state to an insulating state becomes apparent as the nitrogen concentration reaches a threshold of 1/3. This investigation sheds light on the potential of anion engineering in oxide semimetals, offering pathways for manipulating their physical properties. These insights hold promise for future applications that harness these materials for tailored functionalities.

**Main text**

Dirac semimetals emerge as remarkable quantum materials, brimming with enigmatic topological peculiarities. In the Dirac semimetal, the conduction and valence bands contact only at discrete (Dirac) points in the Brillouin zone and disperse linearly along all directions around these critical points [1-4], exhibiting giant diamagnetism, linear quantum magnetoresistance, and quantum spin Hall effect. Researchers have employed many strategies to realize precise control over the structure and physical properties of Dirac semimetals, encompassing the incorporation of 3*d* magnetic elements exhibiting strong correlations through doping schemes [5] and the fabrication of heterostructures incorporating magnetic and/or superconducting materials [6]. However, the investigation pertaining to *in situ* regulation methods remains relatively unexplored, and there is a dearth of comprehensive research in this domain. The study of the physical properties of transition metal oxides reveals its paramount significance in scientific research. Recently, the 3D Dirac semimetals have been experimental explored in oxides, such as *β*-cristobalite $BiO_2$ [1] and $SrNbO_3$. Ok *et al.* [4] reported that $SrNbO_3$ thin film exhibit an emerging topological band structure. By manipulating the octahedral rotations in transition metal oxides, a novel semimetal was created in strained $SrNbO_3$, leading to the realization of a novel correlated topological quantum state. Although this method exhibits commendable efficacy, it is not devoid of limitations, encompassing challenges such as the large degree of stress regulation. Consequently, to facilitate the comprehensive exploration of both *in situ* control and the physical properties of oxide semimetals, novel methodologies need to be adopted.

The burgeoning field of anion engineering in transition metal oxides holds immense promise in fine-tuning their physical properties through the introduction of anions with different size, electronegativity, and charge [7-10]. Heteroanionic materials, exemplified by oxyhydrides [11,12], oxynitrides [13,14], oxyfluorides [15,16], and others [17,18], present a fertile ground for the exploration of novel or enhanced responses, including superconductivity [19], giant magnetoresistance effect [20,21], and visible photocatalytic activity [22,23]. In particular, a technology of *in-situ* nitrogen doping in oxide films is realized through the combination between pulsed laser deposition and a radio frequency nitrogen atom source [24,25]. This approach exhibits simplicity, efficiency, and environmental compatibility, and also enables real-time adjustment of dopant concentration. Consequently, the method is poised to take on a great role in the precise control and regulation of material properties.

In this paper, we report the successful synthesis of highly epitaxial SrNbO$_3$ (SNO) thin films and SrNbO$_{3-\delta}$N$_\delta$ (SNON) oxynitride thin films using *in situ* nitrogen (N) doping. This deliberate substitution of N drives consequential alterations in the valence state of Nb ions and subsequently modifies the electrical transport properties of the materials under investigation. Furthermore, we discuss the strain dependency exemplified by both SNO and SNON films. These observations are in excellent agreement with our first-principles calculations. In this manner, utilizing the paradigm of anion engineering, we unveil a meaningful pathway toward the regulation and manipulation of the distinctive properties characterizing the semimetal phase in oxide thin films.

**Results**

**Structural transition induced by nitrogen doping**. The SNO and SNON thin films with thickness of ~20 nm were grown on (001)-oriented SrTiO$_3$ (STO) substrates by pulsed laser epitaxy. A radio frequency plasma source was used to generate highly active nitrogen atoms for doping N$^{3-}$ ions into the SNO films during the synthesis (see Supplemental Material, experimental section). Figure 1a shows a comparison of XRD $\theta$-$2\theta$ scans from SNO and SNON films. The clear thickness fringes and sharp diffraction peaks indicate the high-quality epitaxial growth for both films. We calculated the out-of-plane lattice constants of SNO and SNON films to be (4.06 ± 0.02) Å and (4.14 ± 0.02) Å, respectively (inset of Figure 1a). The films elongate along the [001] orientation by 2% after N doping. We performed the optical second harmonic generation (SHG) polarimetry measurements on both samples to determine the symmetry variation after N doping. Figures 1b and 1c show $I^{2\omega}_{p-out}$ and $I^{2\omega}_{s-out}$ SHG signals for SNO (blue curves) and SNON (red curves) thin films, respectively. The experimental results were fitted by different models. The best fits to the data were given by the point group symmetry of *mm2* for the undoped SNO films and *4mm* for the SNON films, respectively. These results indicate that N incorporation improves the spatial symmetry of the lattice. Furthermore, we performed scanning transmission electron microscopy (STEM) measurements on a SNON film to illustrate the microscopic structure and chemical distribution. A cross-sectional high-angle dark-field (HAADF) STEM image indicates a chemically sharp and coherent interface between SNON film and STO substrate (Figure 1d and Supplementary Figure S1, [26]). The selected area electron diffraction (SAED) pattern from a SNON film along the pseudocubic [100] zone axis was shown in the inset of Figure 1d. Spatially resolved electron energy loss spectroscopy (EELS) mapping was performed at Nb $L$-, Ti $K$-, N $K$-, and O $K$-edges, respectively, as shown in Figures 1e-1h. The chemical analysis suggests that the N element is uniformly and randomly distributed within the SNON film. To assess the N content and its distribution in the SNON films, we further performed Time-of-Flight Secondary Ion Mass Spectrometry (ToF-SIMS) measurements (Supplementary Figure S2). Using the $NO^-$ signal as a surrogate for N, the depth profiles reveal both a N content enrichment and uniform distribution within the SNON film.

**Electronic states of SNO and SNON films.** The electronic states of SNO and SNON films were examined by X-ray absorption spectroscopy (XAS) and X-ray photoelectron spectroscopy (XPS) at room temperature. Figures 2a and 2b present XAS spectra at N $K$-edges and XPS spectra of Nb $3p$ from SNO and SNON films, respectively. The XAS spectra are normalized to their pre- and post-edges for direct comparison. As shown in Figure 2a, the N $K$-edge XAS of SNON films exhibit three main features positioned at 389.1, 403.0, and 409.5 eV, respectively. Similarly, XPS results in Figure 2b reveal a distinct N 1s peak located at ≈ 396.5 eV for SNON films. Based on previous studies on N-doped perovskite oxides [27,28], this N 1s peak is attributed to the N chemical state associated with N-Nb bonds. The formation of these N-Nb bonds arises from the substitution of nitrogen atoms in lattice oxygen sites. Additionally, we found that the peak shapes of both Nb $3p$ (Figure 2b) and $3d$ (Supplementary Figure S3c) become broadener after N doping, indicating a significant change in the valence state of Nb ions. To quantitatively assess the change in Nb valence resulting from N doping, we performed a fitting analysis on the Nb $3d$ XPS spectra with a Shirley background and three sets of spin-orbit doublets, as shown in Figures 2c and 2d. Specifically, the area ratio between the $3d_{3/2}$ and $3d_{5/2}$ peak pairs was held constant at 2:3, while the spin-orbit splitting energy difference was fixed at 2.75 eV. Based on previous studies [29-32], the intense doublet (purple) at higher binding energy, ~206.8 eV, is assigned to $Nb^{5+}$, while the other two doublets located at lower binding energies, ~206.0 eV and ~204.4 eV, correspond to the $Nb^{4+}$ and $Nb^{2+}$ states, respectively. In the case of stoichiometric SNO, the Nb valence state should be $Nb^{4+}$. The emergence of the $Nb^{5+}$ state can likely be attributed to surface over-oxidation, while the $Nb^{2+}$ state may be associated with the presence of a small amount of oxygen vacancies. Notably, the $Nb^{5+}$ content significantly drops while the $Nb^{4+}$ content increases after N doping, suggesting that N doping could prevent the surface over-oxidation and make SNO more stable.

**Electrical transport properties of SNO and SNON films and their strain dependency.** Transport properties of as-grown films were further examined using standard van der Paw geometry. Figure 3a shows the temperature dependent resistivity ($\rho$) of SNO and SNON films. For thin films grown on STO substrates, the SNO films undergo a clear transition from metallic to semiconducting behavior after N incorporation. The $\rho$ of SNO films increases by three orders of magnitude at room temperature and further enhances to seven orders of magnitude at 10 K after N doping. In addition, Hall measurements confirmed that the majority carriers for both samples are electrons. We plotted the carrier concentration ($n$) as a function of temperature for both films. At room temperature, the $n$ of SNO films is ~$10^{22}$ cm$^{-3}$, consistent with the calculated value based on a $d^1$ electron configuration of $n^{\text{theory}} \approx 1.53 \times 10^{22}$ cm$^{-3}$ [4]. After N doping, the $n$ of SNON films decreases by an order of magnitude compared to that of SNO films. This reduction in $n$ becomes even more obvious, reaching two orders of magnitude at low temperatures. A similar trend was observed in carrier mobility ($\mu$) (Supplementary Figure S5), suggesting the effective modulation of electronic behavior by N doping.

We further investigated the influence of N doping on the electrical transport properties of SNO films at different strain states. The SNO and SNON films were epitaxially grown on (001)-oriented $KTaO_3$ (KTO, $\varepsilon \approx -0.85\%$) and $LaAlO_3$ (LAO, $\varepsilon \approx -5.79\%$) substrates, together with STO ($\varepsilon \approx -2.93\%$). Here, $\varepsilon$ represents the in-plane misfit strain at the interface and is defined by $\varepsilon = (a_s - a_f)/a_s$, where $a_s$ and $a_f$ are the bulk lattice parameters of the substrate and SNO film. The structure characterizations of these strained films were shown in Supplementary Figure S4. All samples show a significant elongation along the out-of-plane direction after N doping. Although the $c/a$ ratio of SNO films reduces dramatically as increasing the compressive strain, the $c/a$ ratio of SNON films exhibits a slight unexpected increase. This abnormal behavior is attributed to the strain relaxation at a high level of compressive strain. The SNO films grown on all these three substrates exhibit a metallic phase across all temperatures we measured. However, when subjected to either a heavy compressive strain or a tensile strain, the room-temperature resistivity increases by an order of magnitude. Similarly, the SNON films on STO and LAO become an insulator at room temperature and show a reduced $n$, whereas the SNON films on KTO undergo an insulator-to-metal-to-insulator transition as decreasing temperature. We fitted $\rho$-T curves of the SNON films using the 3D Mott's variable range hopping (VRH) model, as described by $\rho(T) = \rho_0 + A exp\left(\frac{\Delta E}{k_B T}\right) + B exp(\frac{T_0}{T})^{\frac{1}{4}}]$ formula, where $\Delta E$ describes the activation energy parameter (Supplementary Figure S6) [33]. We obtained $\Delta E \approx$ 23.6, 22.0, and 45.0 meV for SNON films grown on KTO, STO, and LAO substrates, respectively (Figure 3f).

**Band structure and first principles calculations.** To gain deeper insights into the impact of N doping on the electronic structure of SNO films, we show the occupied (Figure 4a) and unoccupied (Figure 4b) electronic states around Fermi level ($E_F$) for both SNO and SNON films, which were taken from XPS valence band (VB) spectra and O $K$-edge XAS, respectively. As shown in Figure 4a, a clear finite density of states (DOS) is observed across $E_F$ for SNO, confirming its metallic behavior (Figure 3). The VB spectrum of SNO, consistent with previous report [31], consists of three features (labeled as A, B, and C) around $E_F$. According to previous studies [31,32], the occupied states close to $E_F$ (feature A) are assigned to Nb 4$d$ states, while feature B is mostly formed by O 2$p$ bands and featured C is assigned to Nb 4$d$/O 2$p$ and Nb 4$p$/O 2$p$ hybridized bands, respectively. After N doping, the vanishing DOS across $E_F$ matches well with the insulating behavior of SNON films. Moreover, a broad shoulder (labeled as feature D) located at the top of feature B appears.

Additionally, O $K$-edge XAS of SNO and SNON films show a consistent change in the spectral line shape and peak intensity, as shown in Figure 4b. The characteristic features at the low-energy region reflect the bonding strength between transition metal cations and ions [34]. The shape resonances in the low-energy region come from two main transitions: (i) from O 1$s$ to hybridized O 2$p$-Nb 4$d$ (Nb-O bond), see peaks in a purple background, and (ii) from O 1$s$ to hybridized O 2$p$-Sr 4$d$ (Sr-O bond), refer to peaks in a green background [35]. Due to the crystal-field interactions, the four shape-resonances reflect the splitting of main transitions (i) (531.5 and 532.6 eV) and (ii) (536.0 and 537.7 eV) into $t_{2g}$ and $e_g$ levels [36]. First of all, the peak intensity of (i) (Nb-O bond) in the SNON film is lower than that in the SNO film, indicating a weaker hybridization between O $2p$ and Nb $4d$ states after the N substitution [37], in consistent with XAS O $K$-edge results. Secondly, the increment of $e_g$ peak intensity at 532.6 eV and simultaneous the reduction of $t_{2g}$ peak intensity at 531.5 eV imply the increased number of $Nb^{4+}$ ($4d^1$) ions in SNON films.

To corroborate the origin of the features observed in XPS VB as well as the changes in O $K$ pre-edge XAS and the corresponding evolution of electronic structure in SNO, we turn to density functional theory (DFT) simulations. Prior to simulating the scenarios that involves N-doping, we first examined the fundamental properties of the undoped bulk SNO. This includes assessing lattice constants and the magnetic moment of the Nb ion, with a specific focus on their relationship to the effective Hubbard parameter $U_{eff}$ ($= U - J$), where $U$ is the Hubbard parameter and $J$ is the exchange interaction. As depicted in Figure S8a, the lattice constants are optimally adjusted at $U_{eff} = 0$ eV, aligning closely with experimental values [38]. The magnetic moment at $U_{eff} = 0$ eV is effectively quenched, in agreement with its non-magnetic fact [39]. Thus, $U_{eff} = 0$ eV is chosen as the default value for the subsequent calculations. Moving forward, we explore the influence of N doping. Initially, we consider a low N concentration of 1/12 (i. e. δ = 1/4), using both the virtual crystal approximation (VCA) method and the substitution of one O atom by one N atom in one unit cell. In the latter case, we examine both equatorial and apical O-sites. For comparison, the DOS of these two cases are present in Figures 4c and 4d, revealing remarkably similar DOS profile although different methods were employed during the simulations. Consequently, it is reasonable to use VAC method in other doping cases, mitigating the need to consider an excessive number of N atom doping configurations. In the VCA method, the SNON system can be effectively simulated with varying N concentration by adjusting the relative weights of N and O atoms. In fact, our experiments did not find any specific ordered arrangement of doped N ions in SNON. Thus, the VCA method is even better in the DFT calculations, than the N-ordered state. The DOS for varying N concentrations are present in Figures 4e-4i. As the N doping concentration increases, there is a gradual upward shift in the conduction bands originating from Nb 4*d* orbitals. In Figure 4j, we illustrate the band diagram schematics for SNO, SNON$_1$ (VCA, 1/6) and SNON$_2$ (VCA, 1/3), with increasing the N doping level. When the N concentration reaches 1/3 (i. e. $\delta$ = 1), a complete insulating state emerges, resulting in full Nb$^{5+}$ (4$d^0$). Further evidence of the metal-insulator transition can be gleaned from the band structures displayed in Figure S9.

**Discussions and conclusions**

In summary, we have unveiled striking structural and transport phase transitions within an semimetallic oxide via effective N-doping. These changes, characterized by substantial alterations in electronic states and carrier density, culminate in a compelling metal-to-insulator transition. This experimental result closely aligns with the predictions made in our theoretical models. Consequently, our study not only offers profound insights into the practical achievement of distinct physical ground states in oxide semimetals through anionic manipulation but also opens up a promising avenue for further exploration and innovation in the realm of advanced materials and condensed matter physics. This work not only deepens our understanding of the intricate interplay between structure and electronic properties but also paves the way for transformative applications in emerging technologies.

**Methods**

**Synthesis of nitrogen doped thin films**

The SrNbO$_3$ (SNO) and SrNbO$_{3-\delta}$N$_\delta$ (SNON) films were grown on (001)-oriented KTO, STO and LAO substrates (Hefei Kejing Materials Technology Co. Ltd) using pulsed laser deposition (PLD). The ceramic target was synthesized by sintering mixtures of stoichiometric amounts of SrCO$_3$ and Nb$_2$O$_5$ powder. The heating process was performed at 20 MPa and 1100 °C for 10 hours. Then, the recovered powder was sintered as a target at 20 MPa and 1000 °C for 12 hours. The undoped SNO films were fabricated in vacuum at the substrate's temperature of 750 °C. The laser furnace was 0.52-0.65 J cm$^{-2}$, and the laser repetition was 2 Hz. For the N-doped SNON films, the highly active nitrogen atoms were in-situ doped into the SNO films during the deposition using RF plasma generated atomic nitrogen under the same experimental conditions. The input power was 250 W and the nitrogen flow was maintained 1.5 sccm/min, keeping the partial pressure of N$_2$ gas at 10$^{-4}$ Torr. The plasma source was equipped with a parallel plate capacitor to remove ionic specie. The films were cooled down to room-temperature under the irradiation of nitrogen plasma in order to compensate the nitrogen vacancies (NVs). The film thickness was controlled by counting the number of laser pulses and further confirmed by X-ray reflectivity (XRR) measurements.

**Structural characterization and elemental mapping**

Synchrotron X-ray diffraction (sXRD) $\theta$-$2\theta$ scans were conducted at the beamline 1W1A of the Beijing Synchrotron Radiation Facility (BSRF). Reciprocal space mapping (RSM) and XRR measurements were carried out using a PANalytical X'Pert$^3$ MRD diffractometer with Cu $K\alpha_1$ radiation equipped with a 3D pixel detector. The thicknesses of films and X-ray scattering length densities (SLD) were obtained by fitting XRR curves using GenX software. Cross-sectional TEM specimen of the SNON/STO film was prepared using the standard focused ion beam (FIB) lift-off process. High-angle annular dark-field (HAADF) images and the selected area electron diffraction (SAED) patterns were taken along a pseudo-cubic $[100]_{pc}$ zone axis using the JEM ARM 200CF microscopy at the Institute of Physics, Chinese Academy of Sciences. Elemental-specific electron-energy-loss-spectroscopy (EELS) and energy dispersive x-ray (EDX) spectroscopy mappings were obtained by integrating the Nb $L$-, Ti $K$-, N $K$- and O $K$-edges signals from selected regions after background subtracting. All data were analyzed using Gatan DigitalMicrograph software.

**Second harmonic generation (SHG) measurements**

Optical SHG measurements of SNO and SNON films on STO substrates were performed in the reflection geometry at room temperature. The pumping beam is a Ti: sapphire femtosecond laser (Tsunami 3941-X1BB, Spectra-Physics, $\lambda$=800 nm). The linearly polarized light incidents on the sample at the angle of 45°. The polarization direction ($\varphi$) of the incident field ($E_\omega$) was rotated through a half-wave ($\lambda/2$) plate. Based on different polarization combinations of incident and reflected light, two configurations were used to conduct the SHG measurement. *P-out* ($I^{2\omega}_{p-out}$) represents the analyzer polarization parallel to the plane of incidence and incident light polarization being rotated, and *s-out* ($I^{2\omega}_{s-out}$) as the analyzer polarization perpendicular to the plane of incidence and incident light polarization being rotated [40,41]. The optical signals were detected by a photon multiplier tube. Theoretical fittings of the SHG polarimetry data were performed with analytical models using standard point group symmetries.

**Electronic state characterization and elemental content**

X-ray absorption spectroscopy (XAS) measurements were conducted for both N *K*- and O *K*-edges at beamline 4B9B of the BSRF. The incident direction of polarized light is parallel to the surface normal. Spectra were collected in the total electron yield mode at ambient temperature. Room-temperature X-ray photoelectron spectroscopy (XPS) measurements were performed at Pacific Northwest National Laboratory (PNNL). Spectra were measured using an electron flood gun to compensate the positive photoemission charge because SNON films were not sufficiently conductive, and conductive SNO films were not grounded. In order to easily visualize the change in Nb 3p and 3d line shapes with N-doping, we align all XPS spectra to place the corresponding O 1s peaks at 530.0 eV. A small polycrystalline Au foil was affixed to the corner of each film surface using Cu tape. For VB spectra, the Au $4f_{7/2}$ peak was used to calibrate the binding-energy scale. The distribution of element valence states was obtained by fitting Nb 3*d* XPS using Casa XPS software. ToF-SIMS measurements were carried out using a ToF-SIMS V (ION-TOF GmbH, Münster, Germany) at PNNL. The mass spectrometer was equipped with a reflection type time-of-flight analyzer. A dual-beam depth profiling strategy was employed, in which a 1.0 keV $Cs^+$ beam (~40 nA, 300 μm × 300 μm scanning area) was used for sputtering and a 25 keV $Bi^+$ beam (~1.1 pA, 100 μm × 100 μm scanning area within the center of the Cs+ crater) was used for negative spectra data collection. Additionally, a flood gun (~1 μA) was used for charge compensation. The film/substrate interface was determined via the secondary ion signals of $NbO^-$ and $^{49}TiO_2^-$.

**Electrical Transport Measurements**

The transport properties were performed by using a 9T-PPMS. All samples were measured in a standard van der Pauw geometry with a four-probe method (to eliminate contact resistance). The contacts were prepared using wire-bonding to ensure that all interfaces were electrically well connected.

**First-Principles Calculations**

First-principles density functional theory (DFT) calculations are performed using the Vienna *ab initio* Simulation Package (VASP) [42,43] based on the projected augmented wave pseudopotentials. For the exchange-correlation functional, the PBEsol (Perdew-Burke-Ernzerhof revised for solids) [44] parametrization of the generalized gradient approximation (GGA) [45-47] method is used . The plane-wave cutoff is set to 500 eV and the Dudarev approach [48] is adopted when Hubbard $U$ is imposed on Nb ion. The atomic positions and lattice constants are fully optimized until the Hellman-Feynman forces converged to less than 0.01 eV/Å. The virtual crystal approximation (VCA) method is used in our calculations which tunes the weight of N and O atoms in one site to simulate the uniform doping [49].

**Data Availability**

The datasets generated during and/or analyses during the current study are available from the first author (H.T.H.) and corresponding authors (E.J.G.) on reasonable request.

**References and notes**

[1] S. M. Young, S. Zaheer, J. C. Y. Teo, C. L. Kane, E. J. Mele, and A. M. Rappe, Physical Review Letters **108**, 140405 (2012).

[2] Z. Wang, Y. Sun, X.-Q. Chen, C. Franchini, G. Xu, H. Weng, X. Dai, and Z. Fang, Phys. Rev. B **85**, 195320 (2012).

[3] Z. K. Liu *et al.*, Science **343**, 864 (2014).

[4] J. M. Ok *et al.*, Science Advances **7**, eabf9631 (2021).

[5] K. Matano, M. Kriener, K. Segawa, Y. Ando, and G.-q. Zheng, Nature Physics **12**, 852 (2016).

[6] H. Boschker and J. Mannhart, Annu. Rev. Condens. Matter Phys. **8**, 145 (2017).

[7] R. D. Shannon and C. T. Prewitt, Acta Crystallographica Section B-Structural Crystallography and Crystal Chemistry **B 25**, 925 (1969).

[8] M. H. Yang, J. Oro-Sole, J. A. Rodgers, A. B. Jorge, A. Fuertes, and J. P. Attfield, Nature Chemistry **3**, 47 (2011).

[9] Y. Kobayashi, Y. Tsujimoto, and H. Kageyama, in *Annual Review of Materials Research, Vol 48*, edited by D. R. Clarke (Annual Reviews, Palo Alto, 2018), pp. 303.

[10] J. K. Harada, N. Charles, K. R. Poeppelmeier, and J. M. Rondinelli, Advanced Materials **31**, 26, 1805295 (2019).

[11] T. Yamamoto, R. Yoshii, G. Bouilly, Y. Kobayashi, K. Fujita, Y. Kususe, Y. Matsushita, K.

Tanaka, and H. Kageyama, Inorganic Chemistry **54**, 1501 (2015).

[12] D. Kutsuzawa, Y. Hirose, A. Chikamatsu, S. Nakao, Y. Watahiki, I. Harayama, D. Sekiba, and T. Hasegawa, Appl. Phys. Lett. **113**, 5, 253104 (2018).

[13] D. Logvinovich, R. Aguiar, R. Robert, M. Trottmann, S. G. Ebbinghaus, A. Reller, and A. Weidenkaff, Journal of Solid State Chemistry **180**, 2649 (2007).

[14] S. G. Ebbinghaus, H. P. Abicht, R. Dronskowski, T. Muller, A. Reller, and A. Weidenkaff, Progress in Solid State Chemistry **37**, 173 (2009).

[15] Y. Kobayashi, M. Tian, M. Eguchi, and T. E. Mallouk, Journal of the American Chemical Society **131**, 9849 (2009).

[16] Y. Tsujimoto, K. Yamaura, and E. Takayama-Muromachi, Applied Sciences-Basel **2**, 206 (2012).

[17] Z. Hiroi, N. Kobayashi, and M. Takano, Nature **371**, 139 (1994).

[18] C. S. Knee and M. T. Weller, Journal of Materials Chemistry **13**, 1507 (2003).

[19] M. Almamouri, P. P. Edwards, C. Greaves, and M. Slaski, Nature **369**, 382 (1994).

[20] A. Kusmartseva, M. Yang, J. Oro-Sole, A. M. Bea, A. Fuertes, and J. P. Attfield, Appl. Phys. Lett. **95**, 3, 022110 (2009).

[21] M. Yang, J. Oro-Sole, A. Kusmartseva, A. Fuertes, and J. P. Attfield, Journal of the American Chemical Society **132**, 4822 (2010).

[22] A. Maegli, S. Yoon, E. Otal, L. Karvonen, P. Mandaliev, and A. Weidenkaff, Journal of Solid State Chemistry **184**, 929 (2011).

[23] B. Siritanaratkul, K. Maeda, T. Hisatomi, and K. Domen, Chemsuschem **4**, 74 (2011).

[24] D. Oka *et al.*, Scientific Reports **4**, 6, 4987 (2014).

[25] D. Oka, Y. Hirose, M. Kaneko, S. Nakao, T. Fukumura, K. Yamashita, and T. Hasegawa, Acs Applied Materials & Interfaces **10**, 35008 (2018).

[26] See Supplemental Material at http://link.aps.org/supplemental/**** for details of further structural, transport, electronic state characterizations and first-principles calculation details.

[27] Y. Y. Mi, S. J. Wang, J. W. Chai, J. S. Pan, C. H. A. Huan, Y. P. Feng, and C. K. Ong, Appl. Phys. Lett. **89**, 3, 231922 (2006).

[28] S. H. Cheung, P. Nachimuthu, A. G. Joly, M. H. Engelhard, M. K. Bowman, and S. A.

Chambers, Surf. Sci. **601**, 1754 (2007).

[29] W. B. Zhang, W. D. Wu, X. M. Wang, X. L. Cheng, D. W. Yan, C. L. Shen, L. P. Peng, Y. Y. Wang, and L. Bai, Surf. Interface Anal. **45**, 1206 (2013).

[30] C. W. Lin, A. Posadas, T. Hadamek, and A. A. Demkov, Phys. Rev. B **92**, 14, 035110 (2015).

[31] C. Bigi *et al.*, Phys. Rev. Mater. **4**, 7, 025006 (2020).

[32] S. Thapa, S. R. Provence, P. T. Gemperline, B. E. Matthews, S. R. Spurgeon, S. L. Battles, S. M. Heald, M. A. Kuroda, and R. B. Comes, APL Mater. **10**, 9, 091112 (2022).

[33] Q. Jin *et al.*, Appl. Phys. Lett. **120**, 073103 (2022).

[34] B. G. Park, Y. H. Jeong, J. H. Park, J. H. Song, J. Y. Kim, H. J. Noh, H. J. Lin, and C. T. Chen, Phys. Rev. B **79**, 035105 (2009).

[35] N. Palina, A. Annadi, T. C. Asmara, C. Diao, X. Yu, M. B. H. Breese, T. Venkatesan, A. Ariando, and A. Rusydi, Physical Chemistry Chemical Physics **18**, 13844 (2016).

[36] A. Chaudhuri, A. Midya, K. Rubi, X. Chi, T. C. Asmara, X. J. Yu, R. Mahendiran, and A. Rusydi, Phys. Rev. B **100**, 085145 (2019).

[37] H. Liu *et al.*, Science **369**, 292 (2020).

[38] H. Hannerz, G. Svensson, S. Y. Istomin, and O. G. D'Yachenko, J. Solid State Chem. **147**, 421 (1999).

[39] A. C. Garcia-Castro, Y. Ma, Z. Romestan, E. Bousquet, C. Cen, and A. H. Romero, **32**, 2107135 (2022).

[40] J.-s. Wang, K.-j. Jin, H.-z. Guo, J.-x. Gu, Q. Wan, X. He, X.-l. Li, X.-l. Xu, and G.-z. Yang, Scientific Reports **6**, 38268 (2016).

[41] Shuai Xu, Jiesu Wang, Pan Chen, Kuijuan Jin, Cheng Ma, Shiyao Wu, Erjia Guo, Chen Ge, Can Wang, Xiulai Xu, Hongbao Yao, Jingyi Wang, Donggang Xie, Xinyan Wang, Kai Chang, Xuedong Bai & Guozhen Yang, Nat. Commun. **14**, 2274 (2023).

[42] G. Kresse and J. Hafner, Phys. Rev. B **47**, 558 (1993).

[43] G. Kresse and J. Furthmüller, Phys. Rev. B **54**, 11169 (1996).

[44] J. P. Perdew, A. Ruzsinszky, G. b. I. Csonka, O. A. Vydrov, G. E. Scuseria, L. A. Constantin, X. Zhou, and K. Burke, Phys. Rev. Lett. **100**, 136406 (2008).

[45] J. P. Perdew, K. Burke, and M. Ernzerhof, Phys. Rev. Lett. **77**, 3865 (1996).

[46] G. Kresse and D. Joubert, Phys. Rev. B **59**, 1758 (1999).

[47] P. E. Blöchl, Phys. Rev. B **50**, 17953 (1994).

[48] S. L. Dudarev, G. A. Botton, S. Y. Savrasov, C. J. Humphreys, and A. P. Sutton, Phys. Rev. B **57**, 1505 (1998).

[49] L. Bellaiche and D. Vanderbilt, Phys. Rev. B **61**, 7877 (2000).

**Acknowledgements**

This work was mainly supported by the National Key Basic Research Program of China (Grant Nos. 2020YFA0309100 and 2019YFA0308500), the National Natural Science Foundation of China (Grant Nos. 11974390, U22A20263, 52250308), the CAS Project for Young Scientists in Basic Research (Grant No. YSBR-084), the China Postdoctoral Science Foundation (Grant No. 2022M723353), the Special Research assistant of Chinese Academy of Sciences, and the Strategic Priority Research Program (B) of the Chinese Academy of Sciences (Grant No. XDB33030200). Synchrotron-based XRD were performed at the beamline 1W1A and XAS /XLD experiments were performed at the beamline 4B9B of the Beijing Synchrotron Radiation Facility (BSRF) via user proposals. XPS and ToF-SIMS measurements, along with the associated data analysis were supported by the U.S. Department of Energy (DOE), Office of Science, Basic Energy Sciences, Division of Materials Sciences and Engineering, Synthesis and Processing Science Program, under Award #10122. ToF-SIMS was performed at the W. R. Wiley Environmental Molecular Sciences Laboratory, a DOE User Facility sponsored by the Office of Biological and Environmental Research. PNNL is a multiprogram national laboratory operated for DOE by Battelle.

**Author contributions**

The nitride samples were grown by H.H. and S.L.; TEM lamellas were fabricated with FIB milling and TEM experiments were performed by T.L. and Q.H.Z.; XPS measurements were performed by J.A.D., B.P., S.A.C., Z.Z., Y.D., and L.W.; H.H., Y.Y.F., D.K.R., Q.J., and S.R.C., worked on the structural and transport measurements. H.Z. and S.D. performed the first-principles calculations based on density functional theory. C.W. and K.J.J. provided important suggestions during the manuscript preparation. E.J.G. initiated the research and supervised the work. L.W., S.D., and E.J.G. wrote the manuscript with inputs from all authors.

**Figures and figure captions**

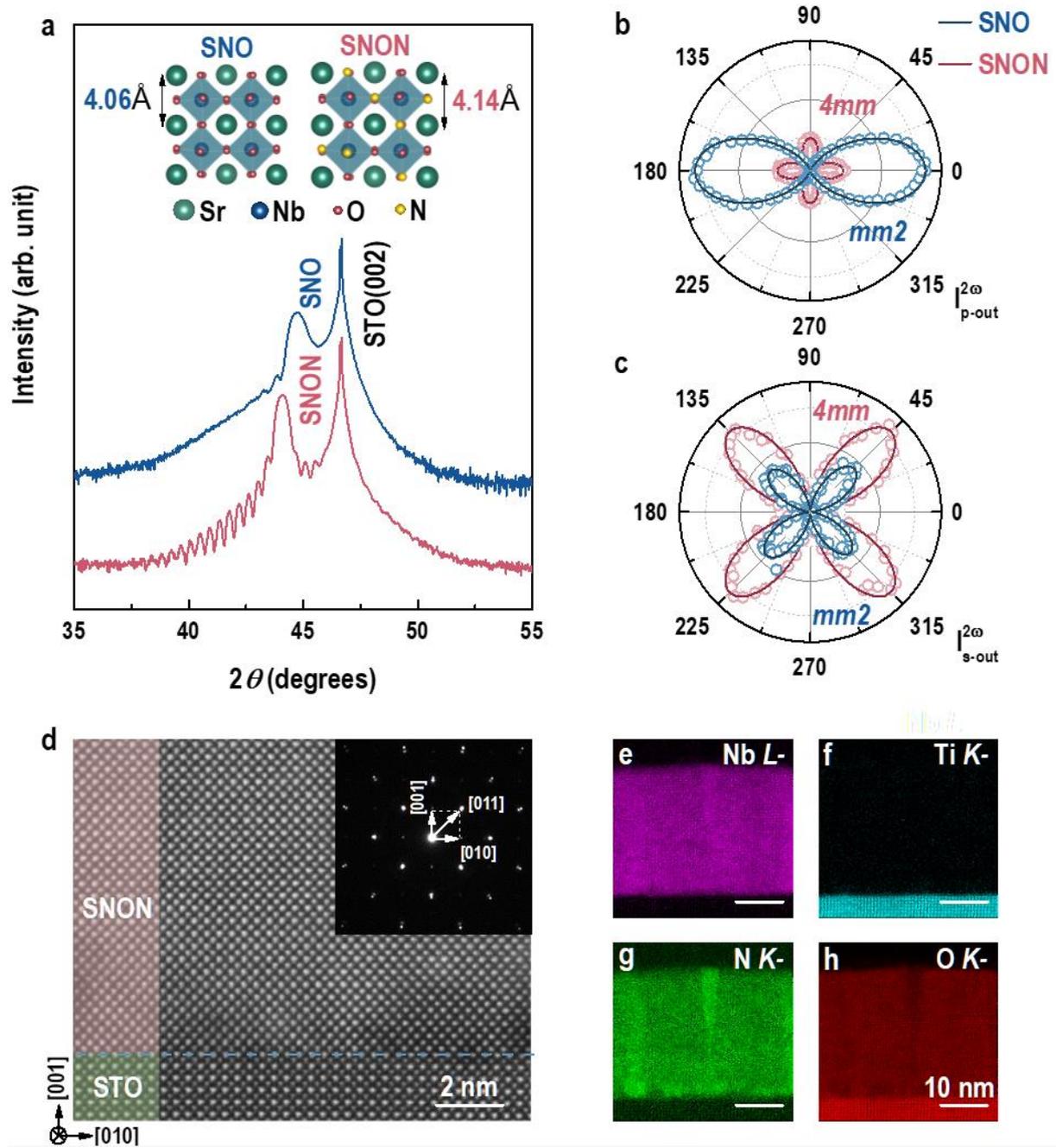

**Figure 1. Structural transition induced by nitrogen doping.** (a) XRD *θ-2θ* scans of SNO and SNON films grown on (001)-oriented STO substrates. Inset: schematic of lattice structures. (b) and (c) SHG signals (open symbols) and theoretical fits (solid lines) for SNO and SNON films, respectively. (d) Cross-sectional HAADF-STEM image and SAED patterns (inset) across SNON/STO interfaces viewed along the [100] orientation. (e)-(h) STEM-EELS maps of a representative region for Nb *L*-, Ti *K*-, N *K*-, and O *K*- edges, respectively.

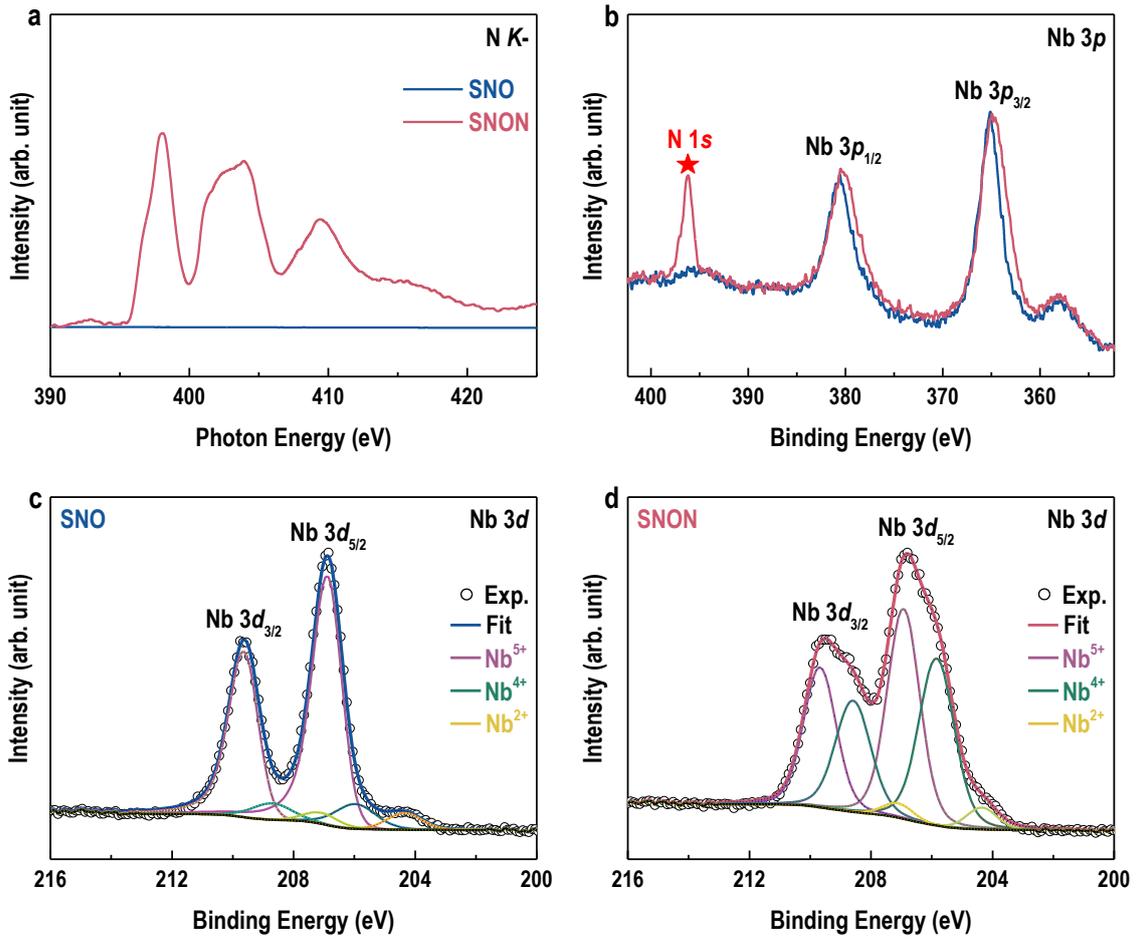

**Figure 2. Electronic states of SNO and SNON films.** (a) N *K*-edges XAS spectra for the SNO and SNON films. (b) N 1*s* and Nb 3*p* XPS spectra for SNO and SNON films. (c) and (d) Measured (open symbols) and fitting (solid lines) results of Nd 3*d* XPS spectra for SNO and SNON films, respectively. The purple, green, and yellow curves are Lorentzian functions fit to the raw spectrum, black curve is the Shirley background, and the blue and red curve are the sum of the individual Lorentzian functions. The area ratio for $Nb^{5+}$, $Nb^{4+}$, and $Nb^{2+}$ is 83.0% (51.0%), 10.6% (44.1%), and 6.4% (4.9%), respectively, for SNO (SNON).

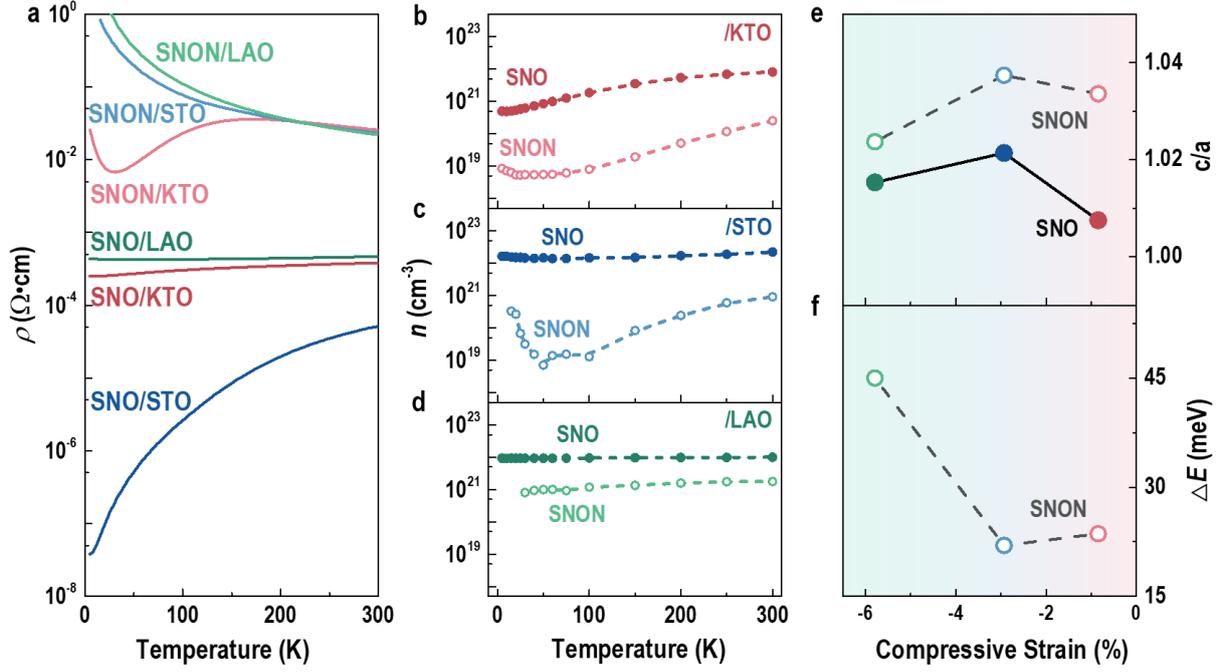

**Figure 3. Strain-dependent transport properties of SNO and SNON films.** (a) $\rho$-$T$ curves of SNO and SNON films grown on KTO, STO and LAO substrates. The temperature dependent carrier density ($n$) of SNO and SNON films on three substrates were shown in (b)-(d). $n$ decreased by 1-2 orders of magnitude after N doping. Strain dependent (e) $c/a$ ratio and (f) activation energy ($\Delta E$) of SNO and SNON films. The impact of the compressive strain on the $c/a$ ratio and $\Delta E$ increases significantly.

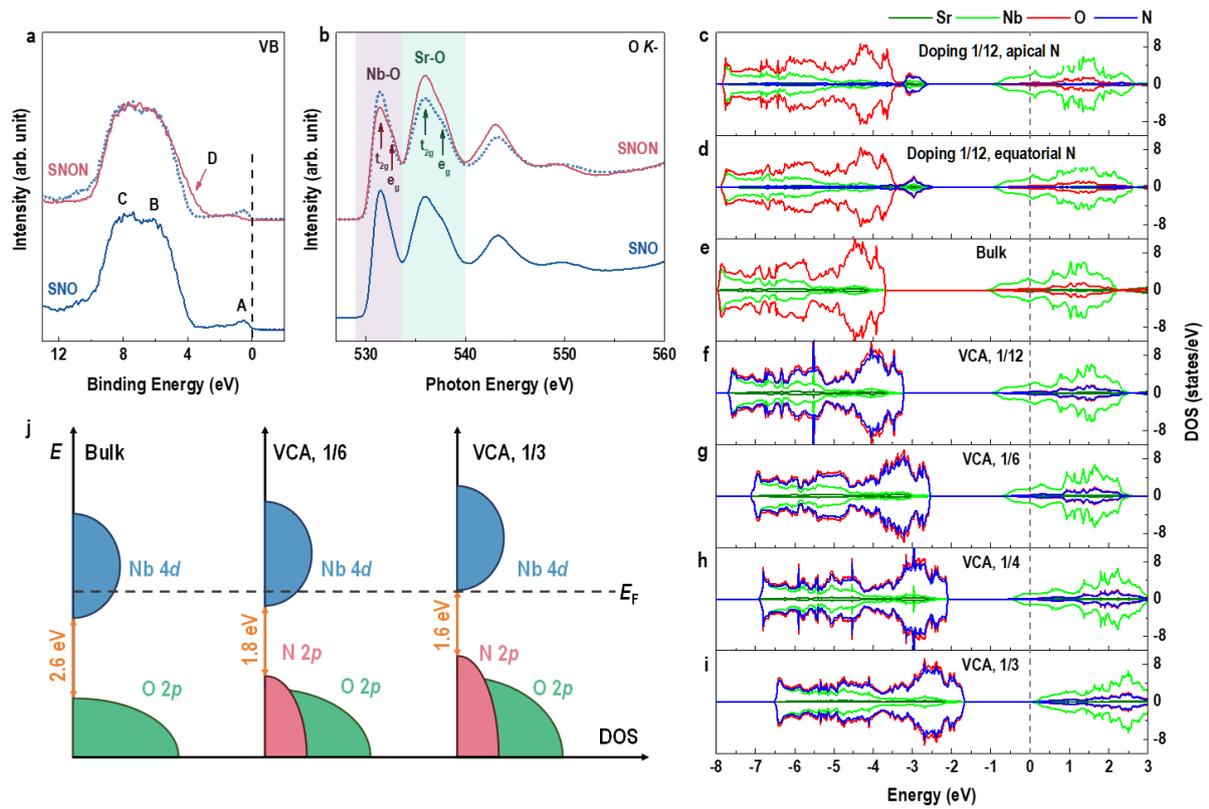

**Figure 4. The evolution of electronic structures upon N doping.** (a) Valence band spectra for SNO (blue) and SNON (red) films. The spectra were shifted to reflect the direct comparison. (b) O $K$-edges XAS spectra for the SNO and SNON films. (c-d) Comparison of DFT DOS for the 1/12 doping case with different methods, which are very similar. (e-i) DFT DOS's around the Fermi level obtained with VCA method for SNO and SNON with N dopant from 0 to 1/3. (j) Band diagram schematics of SNO, SNON$_1$ (VCA, 1/6) and SNON$_2$ (VCA, 1/3), with increasing N doping level.

Supplementary Materials for

# Metal-to-insulator transition in oxide semimetals by anion doping

Haitao Hong,[†] Huimin Zhang,[†] Shan Lin, Jeffrey A. Dhas, Binod Paudel, Shuai Xu, Shengru Chen, Ting Cui, Yiyan Fan, Dongke Rong, Qiao Jin, Zihua Zhu, Yingge Du, Scott A. Chambers, Chen Ge, Can Wang, Qinghua Zhang, Kui-juan Jin, Le Wang,[*] Shuai Dong,[*] and Er-Jia Guo[*]

E-mail: kjjin@iphy.ac.cn, le.wang@pnnl.gov, sdong@seu.edu.cn, and ejguo@iphy.ac.cn

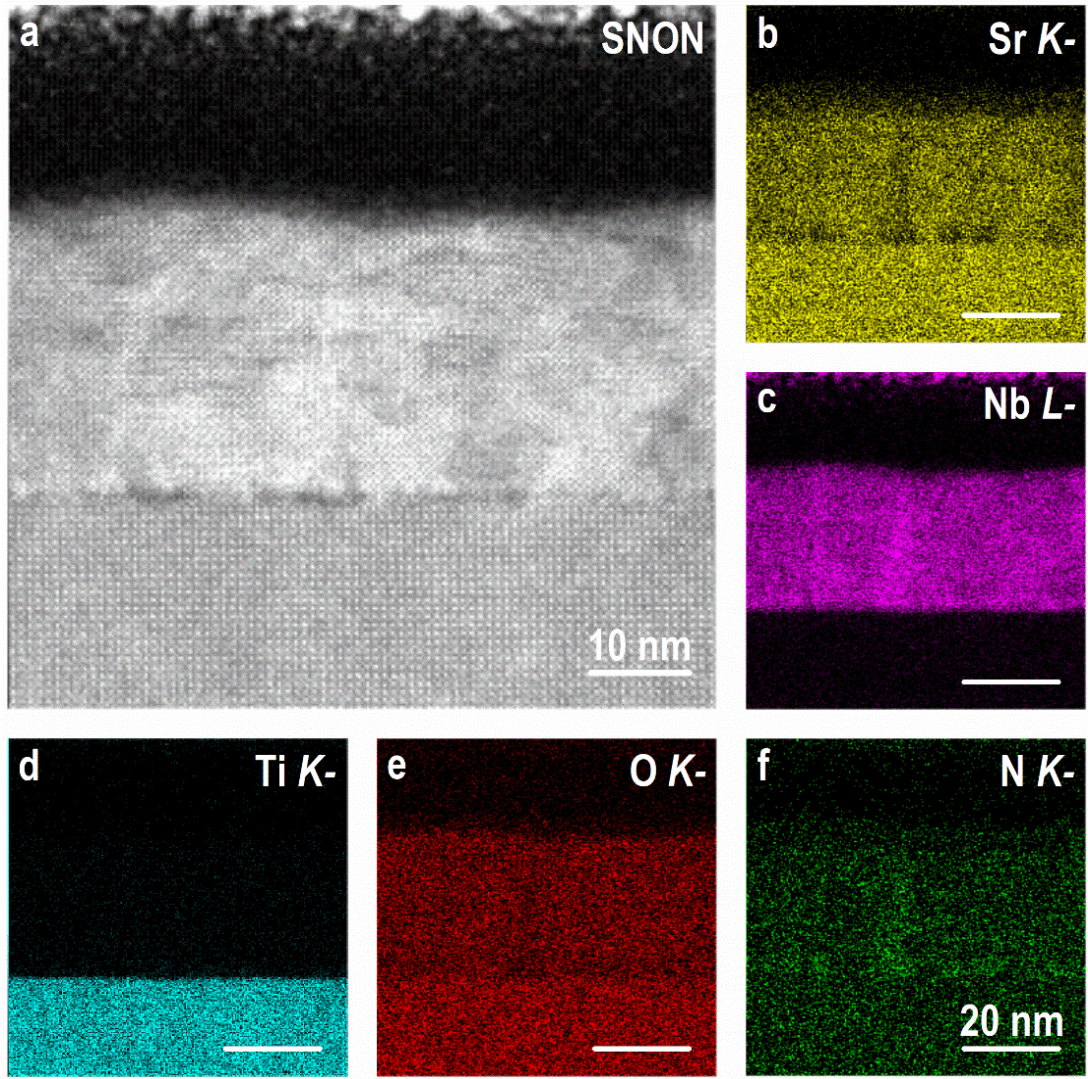

**Figure S1. STEM image and EDX maps of a 23 nm-thick SNON film.** (a) Low-magnified HADDF-STEM image of a SNON film grown on a STO substrate. The STEM results indicate a sharp interface between SNON films and STO substrates. The SNON films show a high crystallinity within the observed region. The colored panels show the integrated EDX intensities of (b) Sr *K*-, (c) Nb *L*-, (d) Ti *K*-, (e) O *K*-, and (f) N *K*-edges. EDS results suggest the elemental distribution is uniform and the SNON/STO interface does not show apparent chemical intermixing. The O content in SNON films is less than that in the STO substrates. We notice that the N content in SNON films is significantly higher than the noise level that present in the STO substrates and Pt coatings.

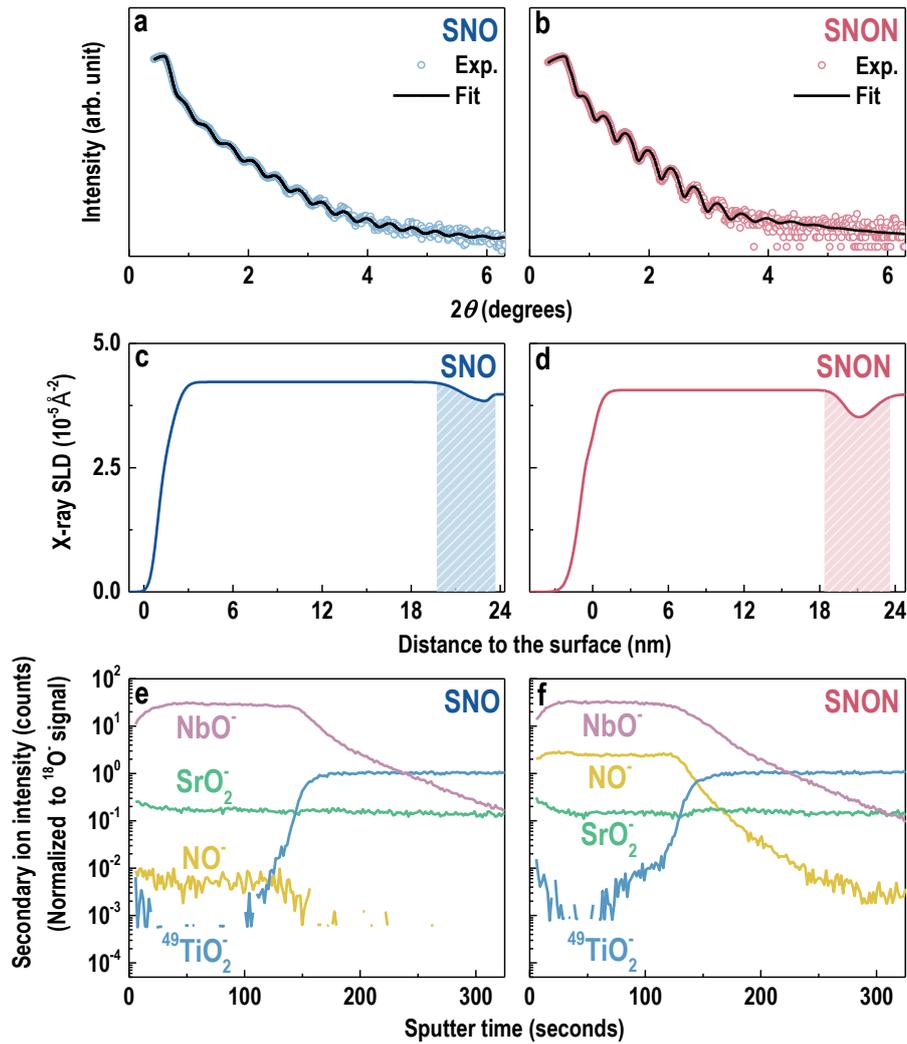

**Figure S2. X-ray reflectometry (XRR) and SIMS measurements on SNO and SNON films.** (a) and (b) XRR curves of SNO and SNON films, respectively. The solid lines are the best fittings to the experimental data (open symbols). (c) and (d) X-ray scattering length densities (SLDs) of SNO and SNON as a function of distance, respectively. At the oxide interfaces, there are unavoidable structural transition and dislocations due to the misfit strain, resulting in a reduction of SLDs at the interfaces (shadow area in both panels). (e) and (f) SIMS results of SNO and SNON films on STO substrates. The interface locations (marked by grey dashed lines) of SNO/STO and SNON/STO were confirmed by the secondary ion signals of NbO$^-$ and $^{49}$TiO$_2^-$. The N signal in SNON is more than two orders of magnitude larger than that in SNO, implying that the N ions were sufficiently doped. The N content in SNO exceeds the natural abundance, possibly attributed to the N introduced during the preparation of the SNO PLD target.

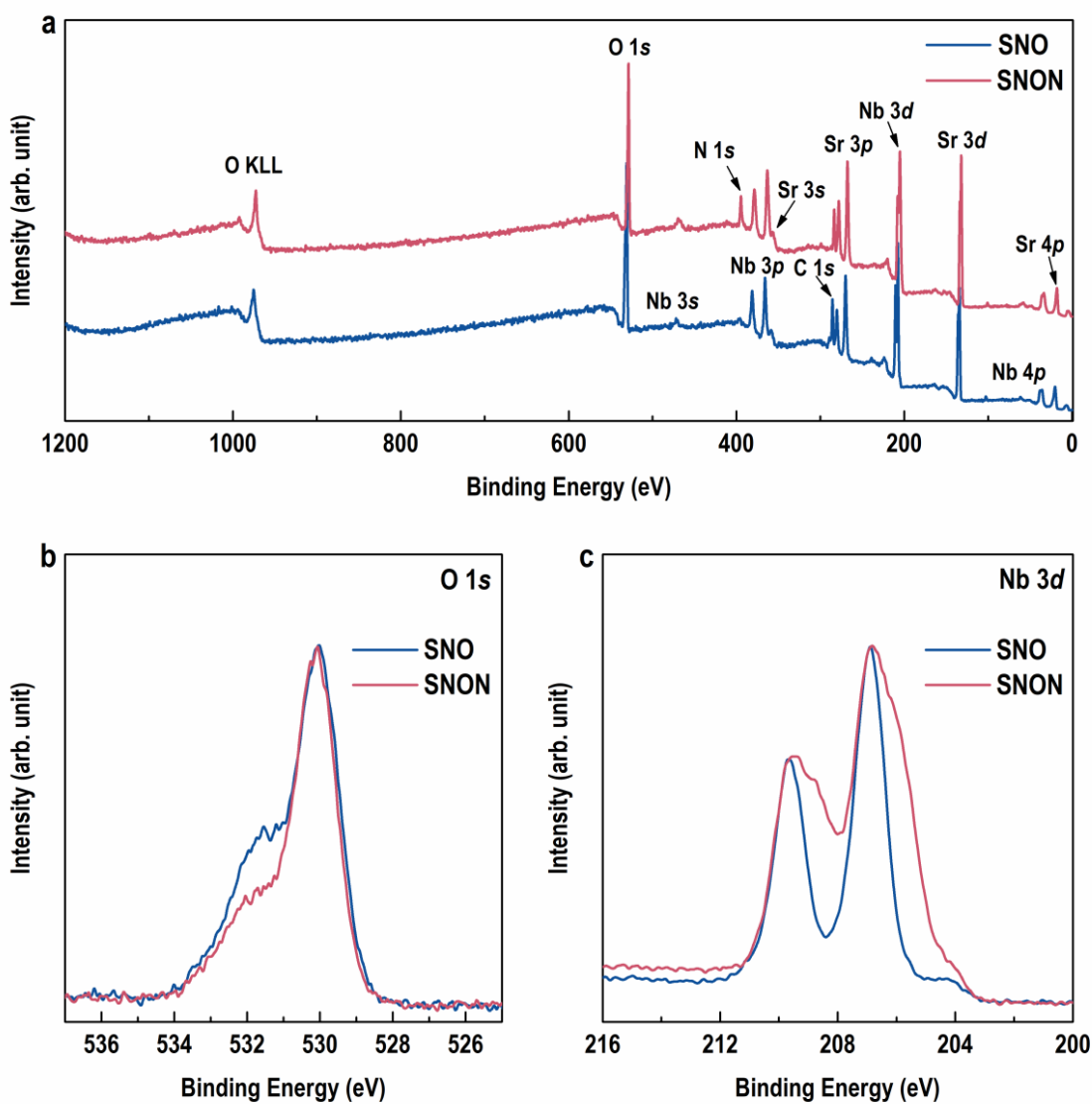

**Figure S3. N-doping increases the Nb valence state.** (a) Survey, (b) O 1$s$, (c) Nb 3$d$ XPS spectra for the SNO (blue) and SNON (red) films. Apparently, a tremendous N 1$s$ signal has been detected after N doping. Compared with SNO, the C 1s signal (a) of SNON displays reduced intensity, resulting in a diminished intensity of the CO3 feature in the O 1s spectrum (b). The existence of Nb$^{5+}$ in the SNO film may potentially arise from surface oxidation processes. We find that the peak intensities of Nb$^{4+}$ increased as well as Nb$^{3+}$ decreased after N inserting, while the Nb$^{5+}$ contribution remains relatively unchanged, which is in consistent with the hole doping.

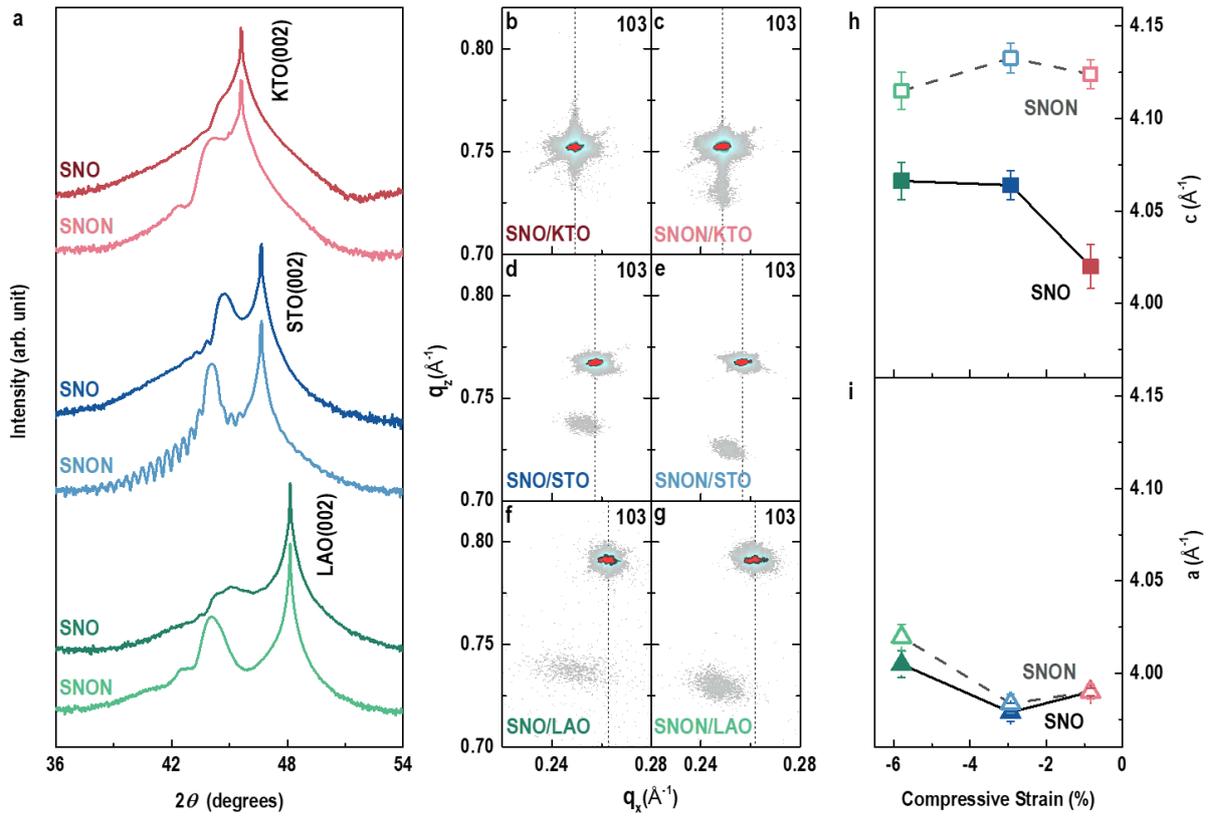

**Figure S4. Structure characterizations of strained SNO and SNON films.** The SNO and SNON films were epitaxially grown on (001)-oriented KTO, STO and LAO substrates. (a) XRD $\theta$-$2\theta$ scans for SNO and SNON films around the substrates' 002 reflection peaks. The film's peaks of strained SNON films shifted to the low angle after N doping, indicating that the *c*-axis lattice constant increases after replacing partial O atoms into N atoms. (b)-(g) Reciprocal space maps (RSMs) around the substrates' pseudocubic 103 reflections for SNO and SNON films grown on different substrates. From RSMs, we could obtain both *a* and *c* lattice constants. As increasing the lattice mismatch, the SNO and SNON films relax their in-plane compression, resulting in a reduction of lattice constants.

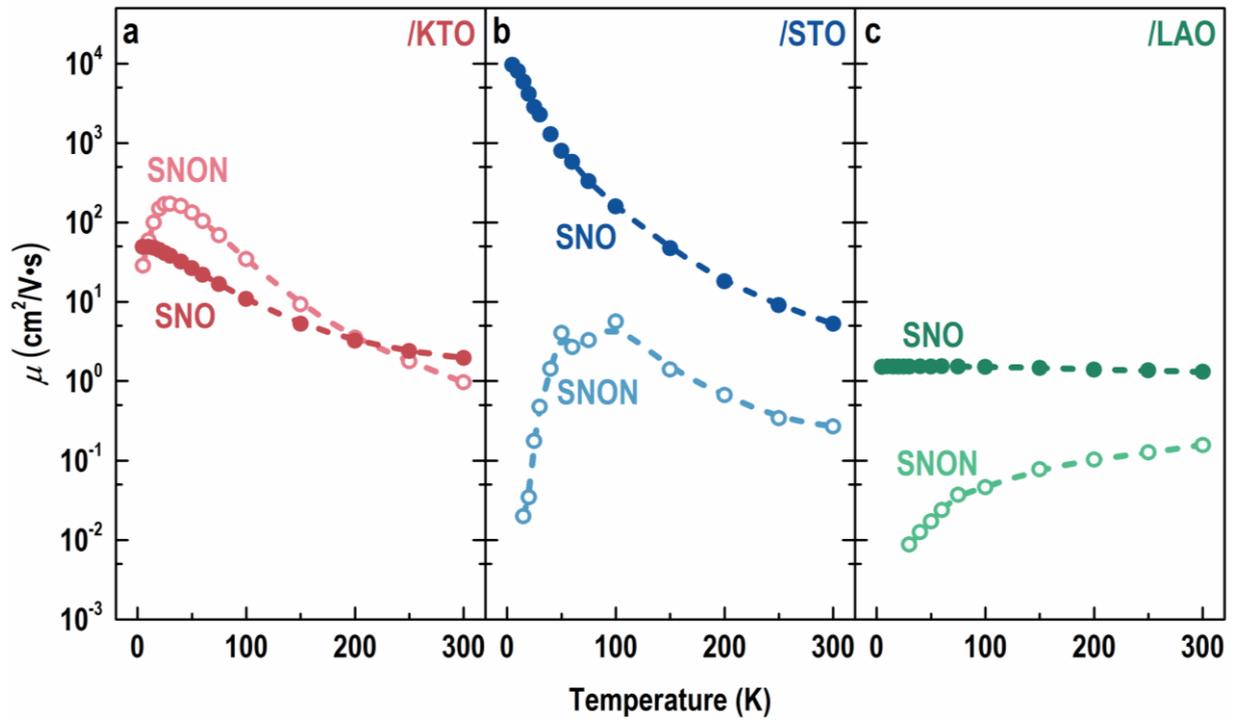

**Figure S5. Carrier mobility (*μ*) of SNO and SNON films grown on different substrates.** (a)-(c) Temperature dependent *μ* for SNO and SNON films grown on KTO, STO, and LAO substrates, respectively. In most cases, *μ* reduces with N doping. With an exception, *μ* of SNON/KTO is slightly larger than that of SNO/KTO at low temperatures.

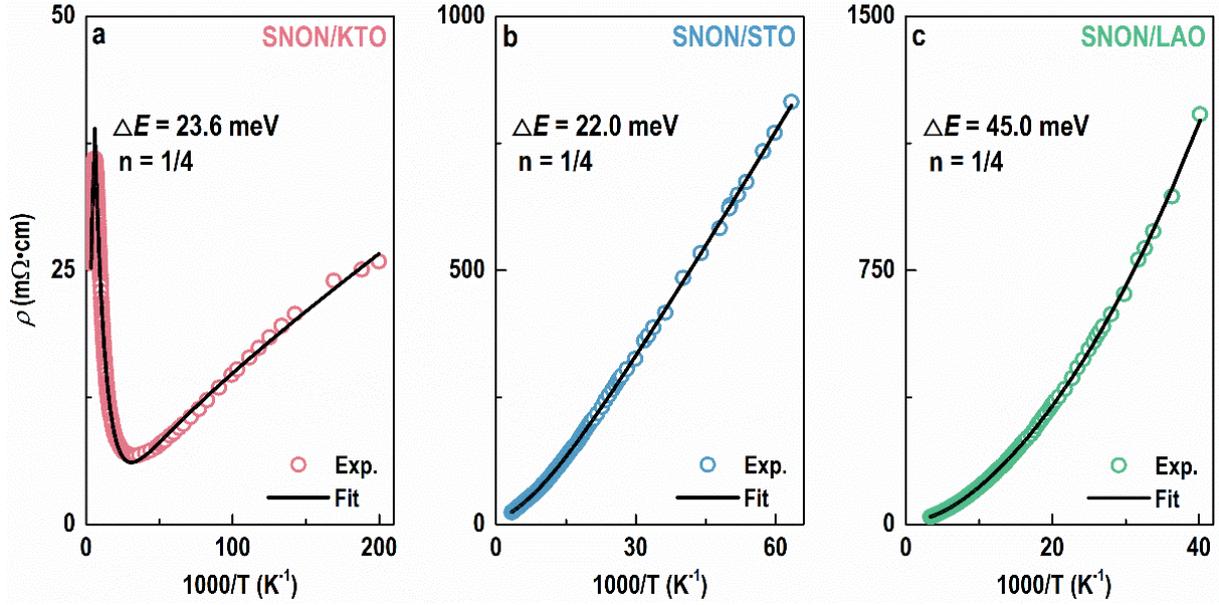

**Figure S6. Fitting curves for transport properties.** (a)-(c) Fitted curves of $\rho$-($1000/T$) for SNO and SNON films grown on KTO, STO and LAO substrates, respectively. All curves fit to the formula: $\rho(T) = \rho_0 + A exp\left(\frac{\Delta E}{k_B T}\right) + B exp(\frac{T_0}{T})^{\frac{1}{4}}]$, where $\rho_0$, $A$, and $B$ are constants and $T_0$ is the temperature normalization constant, respectively. The second term describes the thermal excitation: $\rho(T) \sim \exp\left(\frac{\Delta E}{k_B T}\right)$, where $k_B$ is Boltzmann constant and $\Delta E$ is the activation energy. The third term dominates the Mott's variable range hopping (VRH) model: $\rho(T) \sim \exp\left[\left(\frac{T_0}{T}\right)^n\right]$, where the index $n = 1/4$ is for 3D Mott VRH. The obtained $\Delta E$ at different strain states were plotted in Figure 3 of main text.

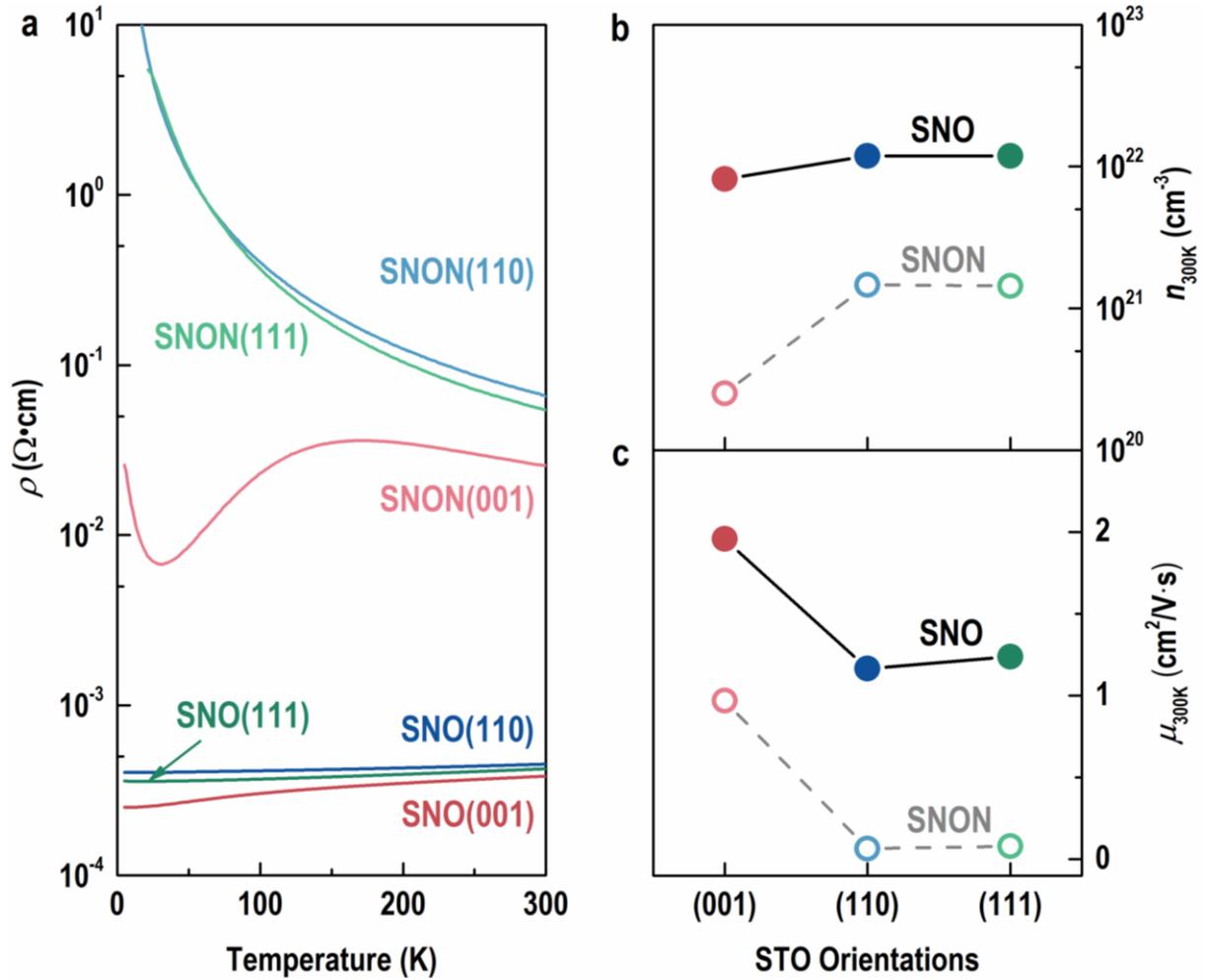

**Figure S7. Transport properties of SNO and SNON films grown on (001)-, (110)-, and (111)-oriented KTO substrates.** (a) $\rho$-$T$ curves of SNO and SNON films. Films grown on different oriented substrates show the same trend: The film gradually changes from the conductivity of SNO films to the insulation of SNON films. (b) and (c) Room-temperature $n$ and $\mu$ for SNO and SNON films, respectively.

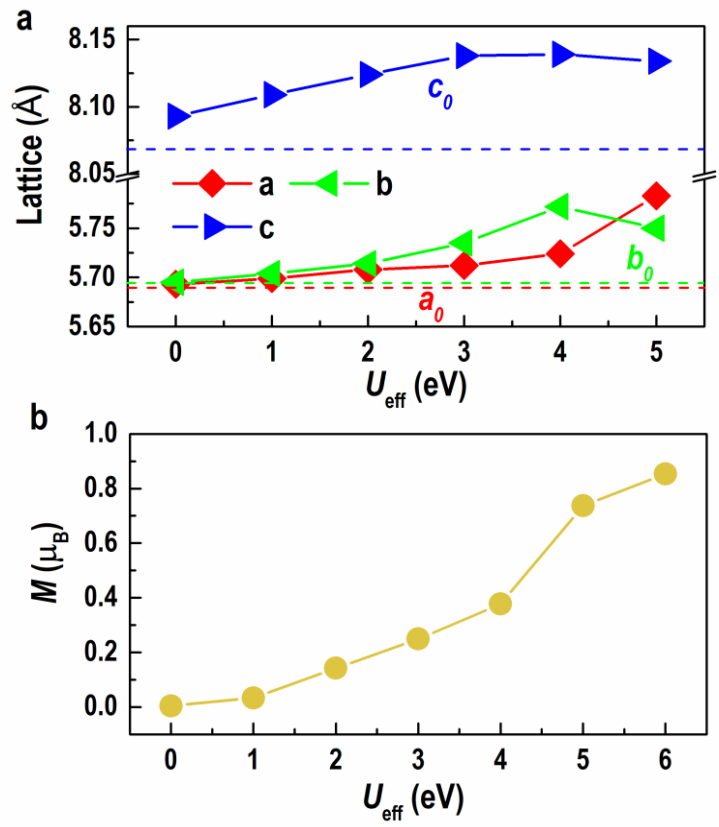

**Figure S8. The tests of $U_{eff}$ value on the lattice constants and magnetic moment of Nb ion.** The dash lines marked with $a_0$, $b_0$, and $c_0$ represent the experimental values [37].

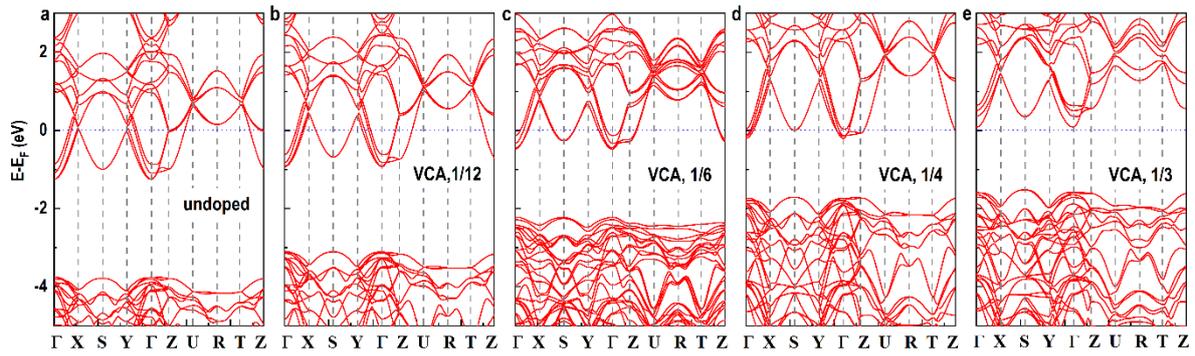

**Figure S9. Band structures of SNON with different N doping levels.** The band structure of SNON changes systematically with N concentrations. Apparently, the conduction bands gradually move upwards across the Fermi level ($E_F$). When the doping level increases to 1/3, the SNON becomes an insulator.